\documentclass[a4,twocolumn]{autart}

\usepackage{subcaption}
\usepackage{makecell}

\usepackage[font=small,labelfont=bf]{caption}

\usepackage[bookmarks=true,hidelinks]{hyperref} 
\usepackage{bookmark}

\usepackage{pdfsync}
\usepackage{graphicx}
\usepackage{amsmath,amssymb}
\usepackage{epstopdf}
\usepackage{amsfonts}
\usepackage{euscript}
\usepackage{mathrsfs}
\usepackage{marvosym}
\usepackage[pdftex]{color}
\usepackage{pdfcolmk}
\usepackage{times}
\usepackage{natbib}  
\usepackage[dvips]{epsfig} 
\DeclareGraphicsExtensions{.ps,.eps}
\newtheorem{example}{Example}

\newcommand{\q}{\quad}
\newcommand{\qq}{\qquad}

\newcommand{\beq}{\begin{equation}}
\newcommand{\eeq}{\end{equation}}
\newcommand{\bea}{\begin{eqnarray}}
\newcommand{\eea}{\end{eqnarray}}

\newcommand{\ese}{\end{subequations}}
\newcommand{\bsea}{\begin{subeqnarray}}
\newcommand{\esea}{\end{subeqnarray}}

\def\Zbb{\mathbb{Z}}
\def\Rbb{\mathbb{R}}

\newcommand{\xb}{\mathbf  x}
\newcommand{\yb}{\mathbf  y}
\newcommand{\zb}{\mathbf  z}
\newcommand{\wb}{\mathbf  w}

\newcommand{\eb}{\mathbf  e}

\newcommand{\ub}{\mathbf  u}

\newcommand{\mub}{\boldsymbol{\mu}}
\newcommand{\varepsb}{\boldsymbol{\varepsilon}}
\newcommand{\varphib}{\boldsymbol{\varphi}}

\newcommand{\xib}{\boldsymbol{\xi}}

\newcommand{\sigmab}{\boldsymbol{\sigma}}

\newcommand{\Phib}{\boldsymbol{\Phi}}

\newcommand{\script}[1]{\EuScript{#1}}
\newcommand{\E}{\mathbb{E}\,}

\DeclareMathOperator{\tr}{\rm Tr\,}

\DeclareMathOperator{\var}{var}

\DeclareMathOperator{\Var} {Var}
\DeclareMathOperator{\Cov} {Cov}

\newcommand{\Frac}[2]{{\displaystyle\frac{#1}{#2}}}

\newcommand{\virg}[1]{``#1''}
\newcommand{\bmat}{\left[ \begin{matrix}}
\newcommand{\emat}{\end{matrix} \right]}

\newcommand{\Ncal}{\mathcal{N}}

\graphicspath{ {./images/} }

\begin{document}

\begin{frontmatter}

\title{\Large An Empirical Bayes approach to\\ ARX Estimation \thanksref{footnoteinfo}}
\author{Timofei Leahu\and Giorgio Picci}
\thanks[footnoteinfo]{T. Leahu was formerly with the School of Mathematical Engineering, University of Padova,Italy {\tt\small timofei.leahu@protonmail.com},\\
 G. Picci is  with the Department of Information Engineering (DEI), University of Padova, Italy,~{\tt\small  picci@dei.unipd.it}}

\begin{abstract}
{\em Empirical Bayes} inference is based on  estimation of  the parameters of an a priori  distribution   from the observed data.  The  estimation technique  of the parameters of the prior,  called {\em hyperparameters}, is based on the {\em marginal} distribution obtained by integrating the joint density of the model with respect to the prior. This is a key step which needs to be properly adapted to the problem at hand.  In this paper we study   Empirical Bayes inference of  linear autoregressive models with inputs (ARX models) for time series and compare the performance  of the marginal parametric estimator with that a full Empirical Bayesian analysis based on the estimated prior.   Such a  comparison, can only make sense for a (realistic) finite   data length.  In this setting, we propose a new  estimation technique of the hyperparameters  by a sequential Bayes procedure which is essentially a backward Kalman filter. It turns out  that for finite data length the marginal  Bayes tends    to behave slightly better than the full Empirical Bayesian parameter estimator and so also in the case of slowly varying random parameters.

\medskip

{\bf keywords: } 
Empirical Bayes Estimation,
Time series Identification,
Conditionally Gaussian ARX models,
Backward conditional  Kalman filtering.
\end{abstract}

\end{frontmatter}

\section{Introduction}\label{IntrEmpBayes}
 The standard estimation technique of AR, ARX or ARMA models  is based on least squares, but this technique, even when  interpretable as Maximum Likelihood, and hence theoretically guaranteeing consistency and asymptotic optimality, requires that the data are generated exactly by an element of  a  rigid model class, therefore assuming existence of true parameters and stationary data. This is often over-optimistic and, even with a reasonable data size one may sometimes obtain  unsatisfactory results. One may hope that describing the model parameters as random variables and using a Bayesian approach could be a reasonable way to improve on  this  problem. However the choice of a prior on the parameter space turns out to be a stumbling difficulty   since in most cases of practical interest no a priori knowledge of a reasonable probabilistic description of the parameters is available. This research is discussing a possible approach to the problem by using an {\em Empirical Bayes} philosophy. Empirical Bayes tries to estimate a prior distribution from the observed data. It looks like  a naive idea and in fact it was proposed  a long time ago see e.g. \cite{Robbins-56} \cite{Casella-85}, \cite{Reinsel-85}. It is only recently that the idea has been revitalized see e.g. \cite{Carlin-L-00},  \cite{Efron-10}, \cite{Efron-14},\cite{Yuanetal-16},  \cite{Petrone-etal-14}, and, due to the availability of massive computer power and suitable algorithms applied to system identification \cite{Aravkin-etal-12}, \cite{Picci-Z-22}.
 It is shown  in the literature that in certain cases Empirical Bayes estimation techniques may even  be superior to Maximum Likelihood \cite{James-Stein-61}, \cite{Efron-M-83}. In our opinion however this claim does not seem to be extendable in sufficient generality as it may depend on various factors, especially on the availability of multiple measurements and the chosen hyperparameter estimation technique.

 In this paper we shall attempt to analyze the  behavior of Empirical Bayes estimation of ARX models by comparing the performance of a  background a priori estimation based on the marginal with the a posteriori estimation   method which uses  the estimated prior within a Bayesian philosophy (which is in fact the Empirical Bayes methodology). We shall not consider comparison with Maximum Likelihood  as its theoretical performance for finite data is  hard to figure   and the asymptotic expression  is considered to be often   too optimistic.\\
    Admittedly ARX is the simplest class of  linear models of time series and its choice is essentially due to simplicity of the analysis. More general or more specific model classes require {\em ad hoc} methods which are outside the scope of this paper. See however \cite{Picci-Z-22} for an example. \\
 Asymptotically, all standard estimation methods for ARX models tend to coincide, and for   a meaningful comparison one should consider {\em finite data} estimates which is obviously the situation one has to deal with in practice. For this reason, we shall first have to review the standard ARX identification material in a {\em conditional setting}. Section \ref{SectARX} will contain a quick review of standard identification  material just to fix notations, and in  the following section \ref{LinModMarg} we shall   proceed to analyze and reinterpret the relative formulas from a {\em conditionally Gaussian perspective}. In sections \ref{CompareEB} and \ref{sec:emp_bayes_estim} we analyze the Mean Square Error (MSE)  performance   of Empirical Bayes estimates and its comparison   with   a  Bayesian estimation method. Finally in Sect. \ref{HypEst} we shall face the critical problem of hyperparameters estimation. This is first addressed  in a step-by step way by extending  some known results on estimation of static linear models but the key final idea is an algorithm  to recover  the (initial) a priori parameters by a  {\em backward Conditionally Gaussian Kalman filter}, derived from scratch in the appendix,  which seems to be new. 
   \section{ARX models and pseudo-linear regression}\label{SectARX}
This is a review section aimed at fixing notations and recalling some standard identification material.\\
{\em Notations:} In this paper, bold   lowercase symbols like $\xb$ denote random quantities, while   italic lowercase letters like $x$ denote deterministic  quantities,  possibly vector-valued with numerical entries (in particular real scalars); the symbol $t$  denotes discrete time running on a certain subset of the integers $\Zbb$; the Mean Square Error (MSE) of an estimator is the sum of the squared bias plus the error variance. Matrix notations are standard, in particular $\tr(A)$ denotes the trace of a square matrix $A$.\\
Imposing an  ARX  structure to  observed output and input  data $\{y(t),\,u(t);\,t=1,2, \ldots,N \}$ means describing them by a linear stochastic difference equation of the form
\begin{equation}\label{ARX1}
 \yb(t)\, = \,\sum_{k=1}^{n} \,a_k\yb(t-k)+ \sum_{k=1}^{m} \,b_k\ub(t-k) + \wb(t)\,, 
 \end{equation} 
where  $\wb:=\{\wb(t),\,t \in \Zbb\}$ is a process of  random errors which is assumed Gaussian uncorrelated (white noise). The data  have been suitably pre-processed e.g. by subtracting the sample mean so as to be compatible with zero-mean and stationarity. Also we shall assume that the input sequence $\{u(t)\}$  is generated independently of the output data $\{y(t)\}$ (no feedback) and can therefore be treated as a deterministic sequence. The ARX model  depends on $p:= n+m$ unknown parameters which will be written as a   column vector:
 \beq\label{Param}
 \theta:= \bmat a_1 & \ldots & a_n & b_1& \ldots& b_m\emat ^{\top}
\eeq
and on the unknown noise variance $\sigma^2$. It  is formally written in {\em pseudo-linear regression} form as
\begin{equation}\label{ARX2}
  \yb(t) =    \varphib(t)^{\top} \theta   + \wb(t)\,,\qq t\in \Zbb
\end{equation}
where $ \varphib(t)$ is a column vector depending on past data:
\beq \label{phi}
 \varphib(t)^{\top}= \bmat  \yb(t-1) & \ldots &  \yb(t-n) & \ub(t-1) & \ldots & \ub(t-m) \emat
\eeq
 Of course there is an important difference with the classical linear regression model, since  the coefficient vectors $ \varphib(t)$ of the model are random, depending on the (past) input-output variables. 
Stacking sequentially the $N$   equations \eqref{ARX2} on top of each other one obtains a system of  relations which, rewritten in  vector form  looks like
\begin{equation}\label{ARX3}
\yb = \Phi_N\, \theta + \wb
\end{equation}
where   the $N$-dimensional random vectors $\yb$ and  $\wb$  have components $ \yb(t)$ and $ \wb(t)$ indexed by   $t=1,2, \ldots,N$ and  $\Phi_N$ is an $N\times p$ matrix of past data of the form:
\beq \label{Phi}
\Phi_N:= \bmat  \varphi(1)^{\top}\\ \vdots \\  \varphi(N)^{\top}\emat \,, 
\eeq
assuming the initial time $t_0$ is far enough, so that we can  fill in 
$\Phi_N$ with the available data so as to describe the output from   time  $t=1$ to $t=N$.   Since   the error process is assumed to be a Gaussian,  one can implement   an {\em exact} Maximum Likelihood procedure to estimate the parameter $ \theta$,  see e.g. \cite[Chap. 8]{Caines-88}, \cite[p.253]{Soderstrom-S-89} or \cite{Hannan-D-88}; however  for finite data  no explicit variance expressions are available. Of course, asymptotically for $N\to \infty$, Maximum Likelihood for stable ARX models is equivalent to an empirical {\em prediction-error minimization} principle  which turns    out to be just  Least Squares, see \cite[p.207]{Soderstrom-S-89} and \cite{Ljung-99}.
 The function of the past data  
\begin{equation}\label{ARXpred}
\hat{ \yb}_{\theta}(t \mid t-1) =    \varphib(t)^{\top}\,  \theta  
\end{equation}
is called the (one step ahead) {\em predictor function} associated to the model. Obviously the predictor function is a linear function of $\theta$ but   is also a function  of the previous $n+m$ past samples of the joint data process.

For an ARX model defined by a generic parameter vector $\theta$,  the  $N$-dimensional vector  of  predictors formally looks like   a linear function of the parameter and the minimization of  the   average squared prediction error $ 
  \varepsb_{\theta}: = \yb - \Phi_N\, \theta
$ 
 leads to the the solution of   a  simple Least Squares Problem. The estimator of  $\theta$ is just
$$
\hat{\theta}_{N}= \left[ \Phi_N^{\top} \Phi_N\right]^{-1}\,\Phi_N^{\top}\,\yb
$$
which is   called a Minimal Prediction Error (PEM) estimator. Asymptotically, assuming the limit for $N\to \infty$ exists, it coincides with Maximum Likelihood \cite{Ljung-99}. It can also be  written in the explicit  form
\begin{equation}\label{ARXSol}
 \hat {\theta}_{N}= \left[\sum_{t=1}^{N}\varphi(t) \varphi(t)^{\top}\right]^{-1}\, \sum_{t=1}^{N}\varphi(t) \yb(t)\,. 
\end{equation}
where one assumes that the inverse exists for suitably large $N$.\\
Clearly $\hat{\theta}_{N}$ is a   non linear  function of the past  observed data so we don't even know when  it may be  unbiased. Even if $\yb$ and $\ub$ were Gaussian, the variance (and indeed the pdf) of $\hat{\theta}_{N}$ for finite sample size would be impossible to compute. One can only  see  what happens for $N \to \infty$. In this paper we shall instead try to do some analysis for {\em finite data length}. To this end we shall have to restrict our goals and  interpret the  formulas as  {\em conditional}  estimates  based on a fixed chunk of observed past data. This   will be the  background idea  for the analysis presented later in this paper.
 
\section{Preliminaries on Conditionally Gaussian Linear Models and Marginals} \label{LinModMarg}
Let ${\mathcal P}_t$ denote the (strict) past observations $\{y(s), u(s); s<t \,\}$ at time $t$. For a Gaussian noise process  we can    interpret the estimator \eqref{ARXSol} as a {\em conditional} Maximum Likelihood estimator given ${\mathcal P}_t$. In fact, under  conditioning with respect to ${\mathcal P}_t$, which in this and in the next  sections we shall maintain fixed throughout, 
 we can assimilate the model \eqref{ARX3} to a 
 {\em conditionally  Gaussian} linear model describing the random vector $\yb$ as depending on a   $p$-dimensional parameter vector $\theta$, written 
\beq\label{Fixeff}
\yb=\Phi_N \theta +\wb
\eeq
where   the entries of $\Phi_N$ (sometimes written   $\Phi$ for short), conditioned on ${\mathcal P}_N$, can be assumed known after the measurements are acquired, and treated as if they were  {\em deterministic}. The error $\wb$ is an unobserved  $N$-dimensional   zero-mean Gaussian vector with uncorrelated components of equal variance $\sigma^2$, which we write  $\wb\sim \Ncal(0, \sigma^2 I_N)$, with $\wb(t)$ independent of ${\mathcal P}_t$ at all times. The model \eqref{Fixeff} is sometimes called a \virg{fixed effects} model since it depends on an unknown but deterministic  parameter vector $\theta$.

 In a  Bayesian setting one describes the parameter $\theta$  as a possible  value taken by a   random vector independent of $\wb$, which will be denoted $\xb$. The model \eqref{Fixeff} with $\xb$ in place of $\theta$ 
 \beq\label{Randeff}
\yb=\Phi_N \xb +\wb
\eeq
  is  called a \virg{random  effects} model.  The  probability density of the random vector $\yb$ in \eqref{Fixeff}, having mean $\Phi\theta $ and variance $\sigma^2I_N$ is now interpreted  as  a {\em conditional density} \footnote{Naturally this density is conditioned also with respect to the previous input-output data ${\mathcal P}_N$ or equivalently $\Phi_N$,  but to simplify notations we shall not indicate this explicitely.}   $p(y \mid \xb=\theta)$, namely
\beq \label{CondTheta}
 p(y \mid \xb=\theta)\equiv{\mathcal N}(\Phi_N \theta, \sigma^2I_N)\,\quad \theta\in \Rbb^p.
\eeq
The random vector $\xb$ is described by some other density, $p_{\xb}(x)$, which we shall also assume Gaussian, in general depending on  unknown mean and variance parameters say
  \beq\label{Prior}
  p_{\xb}\equiv \Ncal(\mu, \sigma^2\Pi)\,.
  \eeq
 Here  for convenience we have introduced a  {\em normalized a priori} variance matrix $\Pi\in \Rbb^{p\times p}$ which we shall assume positive definite. Both $\mu$  and $\sigma^2\Pi$ are unknown.  The {\em joint density} of $\yb$ and $\xb$  is then
\begin{equation}\label{Tomarginal}
p_{\yb,\xb} (y,x)= p(y \mid \xb=x) p_{\xb}(x)
\end{equation}
where   $p(y \mid \xb=x)$ is just the distribution of the original  fixed effects model \eqref{Fixeff} reinterpreted as a conditional density. The {\em marginal  distribution} of $\yb$   could formally be obtained by integrating \eqref{Tomarginal}  with respect to $x$ (of course  this distribution does not depend on $x$ any more). The calculation in terms of density functions  for Gaussian variables can be done in  a straightforward way by just using the expression   \eqref{Randeff}  getting the marginal model for $\yb$  which has now mean $\Phi_N \mu$  and (conditional) variance $\sigma^2(I_N  + \Phi_N \Pi  \Phi_N^{\top})$. In formulas  the marginal density is
 \beq\label{MarginalDistr}
 p_{\yb}\equiv {\mathcal N}(\Phi_N\mu, \sigma^2 R)
 \eeq
 where we have introduced the symbol 
 \beq
 R :=I_N  + \Phi_N\Pi  \Phi_N^{\top}. 
 \eeq
The marginal model \eqref{MarginalDistr} now depends on the unknown parameters $\mu, \sigma^2$ and $\Pi$ which  are  called {\em hyperparmeters} of the distribution. Their estimation could   be approached by applying any estimation principle, say    maximum likelihood, to the  marginal a priori model  \eqref{MarginalDistr}. This would naturally yield an estimate based on the observed data, of the  parametric  prior \eqref{Prior}, in particular of its mean $\mu$ and  variance $\sigma^2\Pi$   which in the Bayesian setting can  take the place of the original fixed-effect model parameters.   According to the terminology  proposed in the literature, e.g.  \cite[p. 262 ]{Lehmann-C-98} and \cite{Carlin-L-00}. Empirical Bayes (EB) inference on  the model \eqref{Fixeff} is really a two step procedure  first recast as inference on  the  marginal   model \eqref{MarginalDistr},   and successively   Bayes a posteriori  inference on the model \eqref{Randeff} using the estimated a priori density. In this paper we have consistently used the  term {\em Marginal} or sometime  the more precise denomination of 
{\em Marginal Bayes} to the estimator of the a priori density based on the marginal  of the output $\yb$ while the term {\em Empirical Bayes} has been  reserved to full Bayesian inference based on the estimated a priori density. 

In general, different estimators, including those based on the fixed effect model could be compared based on their variance and/or mean squared error. The comparison of variances can  in particular   be done using the  {\em conditional expressions} based on past data ${\mathcal P}_N$; in fact letting $V_1({\mathcal P}_N)$ and  $V_2({\mathcal P}_N)$ be conditional variances (possibly matrix-valued) based on the data up to times $N$, we see that, the inequality $V_1({\mathcal P}_N)\geq V_2({\mathcal P}_N)$ (almost surely) implies
$$
\int V_1({\mathcal P}_N)p_N(y,u)dy du\!\geq \!\!\int V_2({\mathcal P}_N)p_N(y,u)dy du
$$
where  $p_N(y,u)$ is the joint distribution of these past data. Therefore optimization of the conditionals implies that of unconditional variances. Hence for comparison purposes we may just work with conditional estimates.
 In this setting  we can  express a conditional  estimate of $R$ given the past observations as an explicit  function of a conditional estimate of the  unknown hyperparameter $\Pi$ and  write the  estimate as:
 \beq
 \hat R =   I_N  + \Phi_N \hat \Pi  \Phi_N^{\top} \,  \label{MLb}
\eeq
 where the estimator $\hat \Pi$ of the prior variance and $\hat\sigma^2$ will be discussed later. All matrices in formula \eqref{MLb} are functions of the observations $\{y(s), u(s); s<N ~\}$ for some   fixed sample of size $N$ while, in particular,  the symbol $\yb$ is used  for the random vector obtained by listing the components of the observed output data updated by one time index,   available at time $N$, say $\{y(s);\, s=1,\ldots, N\}$. 
  
 Naturally  a  preliminary question which should be addressed at this point is the {\em identifiability} of $\Pi$ in the model \eqref{MarginalDistr}. This requires to check the unique dependence (injectivity) of the variance $\sigma^2(I_N  + \Phi_N \Pi  \Phi_N^{\top})$ on the matrix variable $\Pi$ which in turn is equivalent to checking whether the equation $\Phi_N\Delta  \Phi_N^{\top}=0$ has the unique solution $\Delta=0$. This will clearly  be true if the sample size is large enough to make $ \Phi_N $ of full rank $p$ almost surely, an assumption which guarantees identifiability of $\theta$ as well and will be assumed to hold all through. We should however warn the reader that this is a {\em finite sample} requirement. For sample size tending to infinity  $\Pi$ becomes non-identifiable. We shall comment on this later on in Sect.   \ref{HypEst}.
 
\section{Relations between the marginal and Empirical  Bayes Estimates}\label{CompareEB}
 We  first discuss  the estimation of the parameter $\mu$ in the conditionally Gaussian  marginal model \eqref{MarginalDistr}. Assuming for the moment that $\sigma^2$ and $\Pi$ have been estimated somehow and are known so that   the matrix  variance of $\yb$ is  as in  \eqref{MarginalDistr}, then the (conditional)   estimator say $\hat \mu$ of $\mu$ is given by the standard least squares  expression
 \beq \label{EBformula}
 \hat \mub=  \left[ \Phi_N^{\top}R^{-1}\Phi _N\right]^{-1}\Phi _N^{\top} R^{-1} \,\yb
 \eeq
and its conditional variance is given, again by  standard least squares theory, by
 \beq \label{VarEB}
 \Var [\hat\mub]= \sigma^2 \left(\Phi _N^{\top}R^{-1}\Phi _N\right)^{-1} \,.
 \eeq
 To relate the marginal  formula \eqref{EBformula}   to the Bayesian (a posteriori) estimate $\hat \xb= \E[\xb\mid \yb]$  of the  variable $\xb$ in the model \eqref{Randeff},   assuming again that the a priori parameters $\sigma^2, \Pi$ have been estimated and are known,  one  should rely on the standard formulas valid for the linear Bayesian model \eqref{Randeff}. After cancellation of the common factor $\sigma^2$, these   formulas become
 \begin{align}
\hat{\xb}&= \mu+ \Phi _N{\top}\Phi _N + \Pi^{-1})^{-1}\Phi _N^{\top}(\yb -\Phi _N \mu)\label{BayesEst}\\
\Var [\xb\mid \yb]&= \Var (\xb-\hat\xb) = \sigma^2 (\Phi _N^{\top}\Phi _N + \Pi^{-1})^{-1} \label{BayesVar}
\end{align}
where   $\Pi$ is assumed invertible. For next use in the MSE formula, note that \eqref{BayesVar} , i.e. the  conditional variance of $\xb$ given $\yb$, is also the
normalized {\bf error} variance  of the Bayes estimator $\hat\xb$ as it  follows  from the matrix inversion lemma.\footnote{The lemma states that 
$$
\left[ A +BCD\right] ^{-1}= A^{-1} -A^{-1}B\,[ C^{-1} + DA^{-1}B\,]^{-1}DA^{-1}
$$ 
assuming that all indicated inverses exist.} 

It turns out that the Bayesian estimator $\hat \xb$ and the  marginal  estimator $\hat \mub$ have a different  variance.
  \begin{prop}\label{EBVar}
 The variance of the conditional   estimator of $\mu$  based on the marginal  model  \eqref{MarginalDistr}   has the equivalent expressions 
   \begin{equation}\label{varTheta}
\Var [\hat\mub]= \sigma^2 \left( \Phi_{N}^{\top}R^{-1}\Phi_{N} \right)^{-1} =\sigma^2[(\Phi_{N}^{\top}\Phi_{N} )^{-1}+\Pi]\,.
 \end{equation}
\end{prop}
\begin{pf} Follows from the identity
 $$
 R^{-1}=[ I  + \Phi \Pi  \Phi^{\top}]^{-1}= I-\Phi[\Pi^{-1} +\Phi^{\top}\Phi]^{-1}\Phi^{\top}
 $$
 which, again by the matrix inversion lemma gives
  \begin{align}
  \Phi^{\top}R^{-1}\Phi  &  = \Phi^{\top}\Phi-\Phi^{\top}\Phi [\Phi^{\top}\Phi +\Pi^{-1}]^{-1}\Phi^{\top}\Phi = \notag \\
  &= [(\Phi^{\top}\Phi)^{-1} +\Pi]^{-1} \,.
  \end {align}
  \end{pf}
   From \eqref{varTheta}  $\hat \xb$ and $\hat \mub$ are seen to be in general different estimators. 
Moreover while $\hat \mub$ is unbiased, we have instead   $\E\,[\hat \xb\mid \xb=\theta] \neq \theta$ for any $\theta$ and  so $\hat \xb$ is always biased.\\
Note that the variance of the estimator of $\mu$ based on the marginal model is the sum of the variance of the parameter in the classical fixed effect model plus the variance of the prior of $\xb$. Clearly when $\Pi\to 0$ the variance of the marginal estimator tends to the classical fixed effect one while the variance  \eqref{BayesVar} of the Bayes estimator  tends to zero since the parameter becoms a {\em known} deterministic quantity.

\section{The MSE of the Empirical Bayes estimators} \label{sec:emp_bayes_estim}
 
\begin{figure}
    \centering
    \includegraphics[width=.8\linewidth]{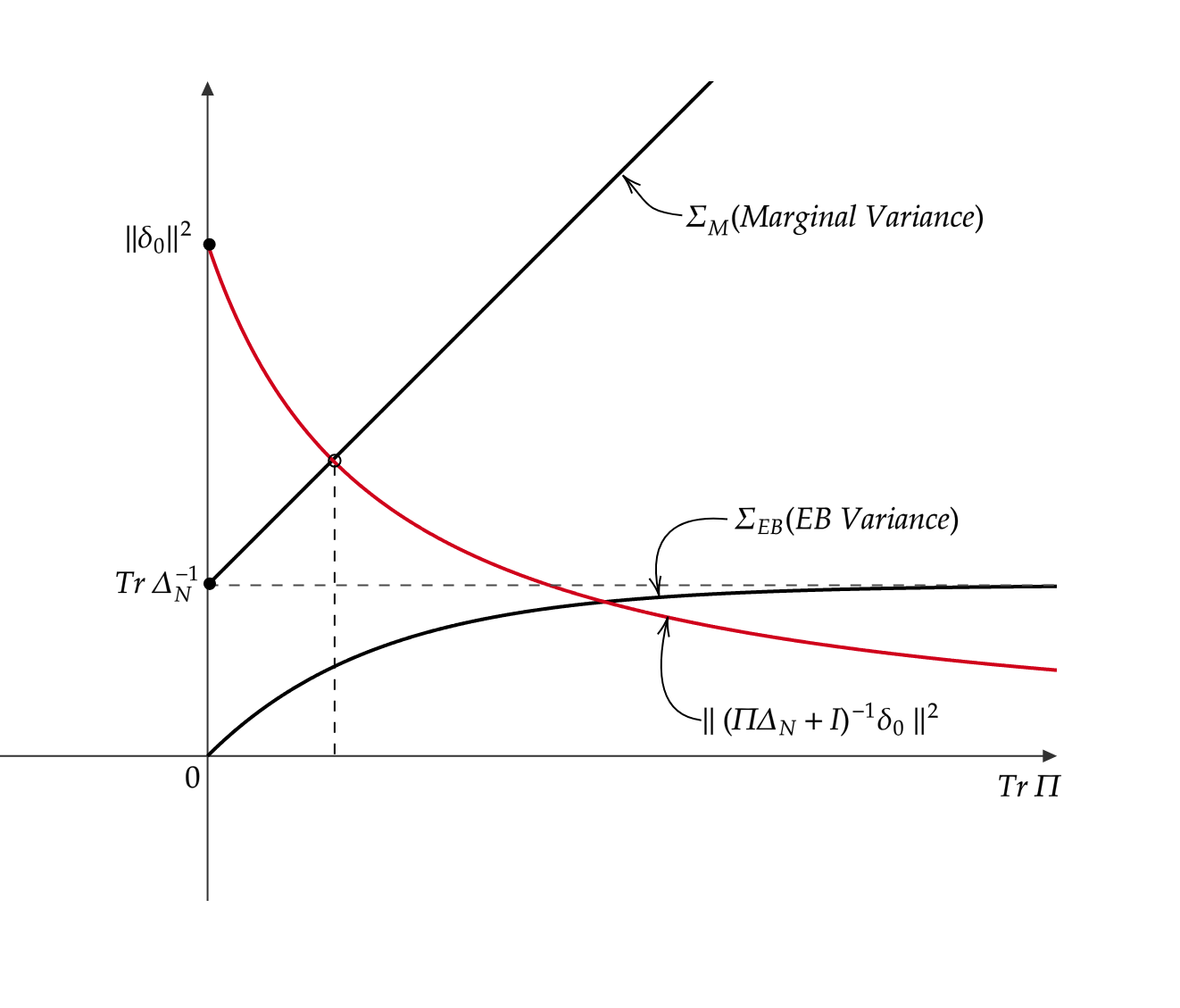}
    \caption{Behavior of the two variances and the bias norm squared (in red).}
    \label{fig:anal-bias-variances}
\end{figure}

In this section we address the comparison of  the  marginal a priori with  the  Empirical Bayes (a posteriori)  estimators of ARX models for a finite sample size. The comparison is based on  the mean-square errors (MSE) of  the two procedures.To this end  we shall use the explicit formulas for the conditional mean and variance of the two estimators  assuming that the past data  have been observed, that is, conditioning everything on past data also in the Bayesian  formulas \eqref{BayesEst}, \eqref{BayesVar}. \\
 To address the comparison we first need expressions for the MSE of the two estimators. The Lemma below is instrumental for this calculation.
\begin{lem}
 Assume that the  normalized a priori variance $\Pi$   of $\xb$  is  positive definite and let $\Delta_N:= \Phi_N^{\top}\Phi_N$ and let  $\E_{\theta_0}$ be the conditional expectation with respect to the event $\xb=\theta_0$, then the bias of the Empirical Bayes a posteriori estimator, defined as $ \E_{\theta_0}(\hat\xb) -\theta_0$ has the expression 
\beq \label{BiasBayes}
\E_{\theta_0}(\hat\xb) -\theta_0= [\Delta_N+\Pi^{-1}]^{-1}\Pi^{-1}(\mu-\theta_0)\,.
\eeq
\end{lem}
\begin{pf}
Follows from \eqref{BayesEst} since $\E_{\theta_0} \yb= \Phi \theta_0$ and hence $\E_{\theta_0}(\hat\xb)-\theta_0=(\mu -\theta_0)-   [\Delta_N+\Pi^{-1}]^{-1}\Delta_N (\mu -\theta_0)$ and then using the identity
\begin{align*} 
&[\Delta_N+\Pi^{-1}]^{-1}\Delta_N  - \Delta_N^{-1}\Delta_N = \\&[\Delta_N+\Pi^{-1}]^{-1}[ I_N-(\Delta_N+\Pi^{-1})\Delta_N^{-1}] \Delta_N \\
&=- [\Delta_N+\Pi^{-1}]^{-1}\Pi^{-1}. 
\end{align*} 
\end{pf}
We shall hereafter denote the error  on the mean $\mu-\theta_0$  by $\delta_0$  and then display the formulas for the theoretical Mean Square Errors of the two estimates as follows.
\begin{prop}
The (scalar) MSE of the Empirical Bayes a posteriori estimator is
\begin{align}
\|\E_{\theta_0}(\hat\xb) -\theta_0\|^2 + \tr\Var(\hat\xb) &= 
\|[\Delta_N+\Pi^{-1}]^{-1}\Pi^{-1}\delta_0\|^2 \notag \\
 &+\sigma^2 \tr[\Delta_N+\Pi^{-1}]^{-1}
\end{align}
while  the conditional  marginal Bayes estimator is unbiased and its MSE coincides with the variance \eqref{varTheta} which can be written as $\sigma^2 \tr[ \Delta_N^{-1}+\Pi] $.
\end{prop}
The first formula follows from \eqref{BayesVar}, \eqref{BiasBayes}. Note that both formulas assume that the hyperparameter $\Pi$ is  known which of course is never the case. According to the underlying   EB philosophy, later on we shall  have to substitute the theoretical  value with   suitable  estimate. See Sect. \ref{HypEst}.\\
To get some intuition on the MSE  formula let us first consider the case in which $\delta_0=0$ (which seems to play the role of  unbiasedness, but one should recall that the Bayes estimator is never unbiased) in which case we just compare variances. In this special case  it   is clear that  the marginal  is  always worse than the Empirical Bayes estimator since for $\Pi \to 0$
$$
\Delta_N +\Pi^{-1} \geq (\Delta_N^{-1} +\Pi)^{-1} \simeq \Delta_N \,.
$$
Moreover for $\Pi$ large the marginal variance $\Delta_N^{-1} +\Pi$ diverges linearly  while that of the Empirical Bayes estimator remains bounded above by $\Delta_N^{-1}$. Compare Fig. \ref{fig:anal-bias-variances}.
 However since any  Bayesian estimator is never unbiased, we need to analyze the effect of the squared bias term and this will make a difference.\\
Let $\theta_0$ be the true parameter which has generated the data according to the fixed effect model \eqref{Fixeff}. Note that for small $\Pi \to 0$ the bias term reduces to $\|\delta_0\|^2$ which could be much larger than $\tr\Delta_N^{-1}$ especially if $N$ is large enough. Therefore if $\|\delta_0\|^2 > \tr \Delta_N^{-1}$ there is room for the MSE of the Empirical Bayes estimate to be larger than that of the Marginal. This  is shown graphically in the Figure \ref{fig:anal-bias-variances}. See the red line in the picture.

\begin{figure}
    \centering
    \includegraphics[width=.8\linewidth]{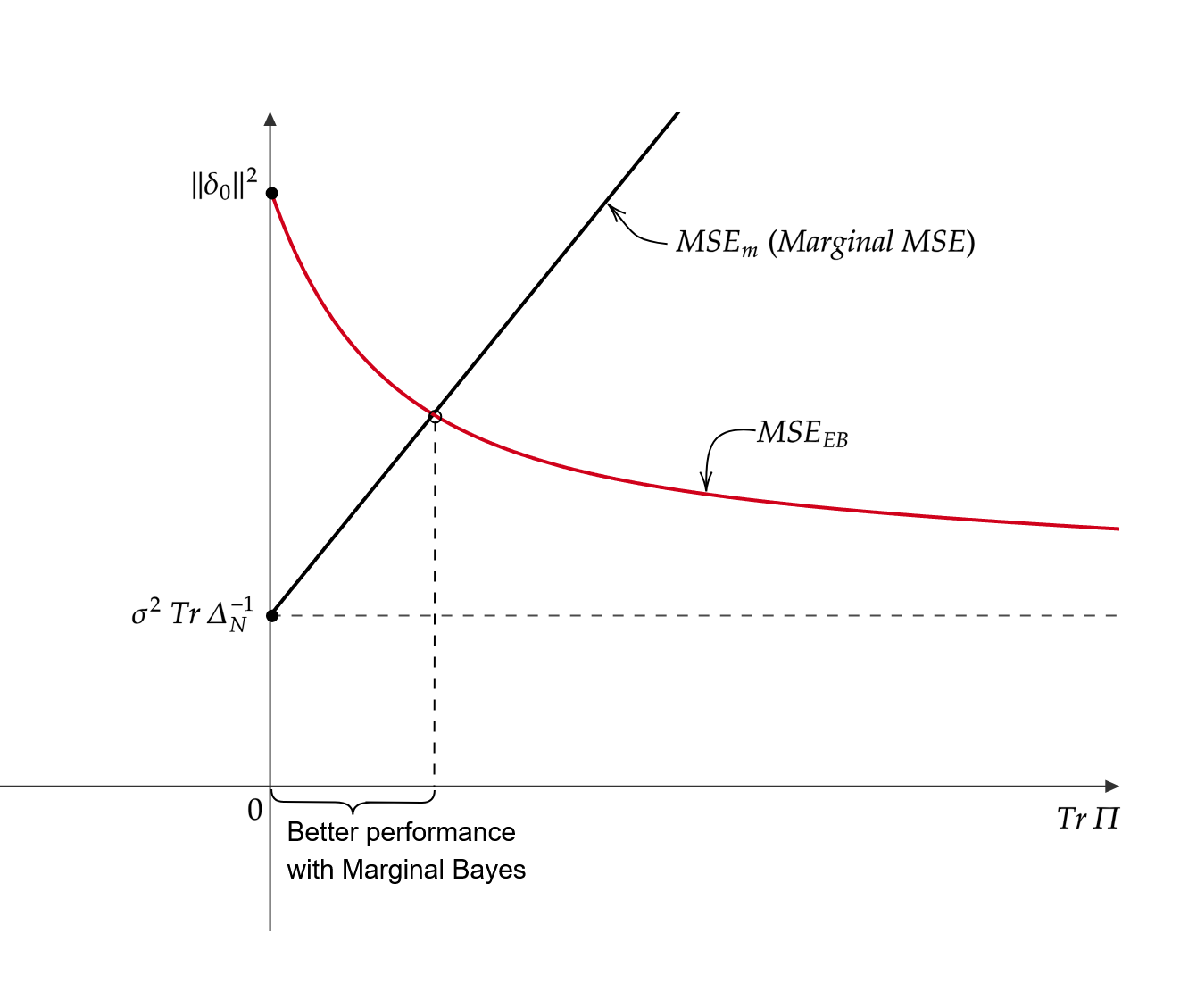}
    \caption{Comparison between Marginal  and Empirical Bayes MSE.}
    \label{fig:compare-EB-Bayes-MSEs}
\end{figure}

 The comparison between the two scalar MSE's is shown in  Figure \ref{fig:compare-EB-Bayes-MSEs}. The important part to be appreciated in this figure is that there is always a non negligible interval in which the marginal estimator gives better performance. \\
  Morale: the simpler marginal estimator  in  circumstances where the a priori variance $\Pi$ is small can give better performance than the full Empirical Bayes. This agrees with intuition as a small variance indicates a "nearly deterministic" parameter.

\begin{example}\label{ex5.1}
{\em Consider the case of a stationary AR model with a scalar parameter $\theta$:
$$
\yb(t)=\theta \yb(t-1) +\wb(t),\qq t=1,\ldots, N
$$
 where the noise has unit variance, the true parameter is denoted $\theta_0$ with $|\theta_0|<1$ and we assume that the a priori mean is $\mu=0$. The normalized a priori variance  is  denoted $\pi\geq 0$ and  $\Delta_N$ is now a positive scalar quantity. The normalized  mean square error of the Empirical Bayes estimate is
\begin{align*}
&e^2_{EB}(N):= \left( \Frac{\pi^{-1}}{\Delta^{2}_N+\pi^{-1}}\right)^2 \theta_0^2 + \Frac{1}{\Delta^{2}_N+\pi^{-1}}\\
&=\Frac{\theta_0^2+\pi+\Delta^{2}_N \pi^{2}}{1+2\Delta^{2}_N \pi+\Delta^{4}_N \pi^{2}},
\end{align*}
which is to be compared with the normalized  mean square error of the marginal say, $e^2_{M}(N)=1/\Delta^{2}_N +\pi$.  

Let us first analyze the two errors in function of the prior $\pi$. For $\pi\to 0$, $e^2_{EB}(N) \to \theta_0^2$ while  $e^2_{M}(N)\to 1/\Delta^{2}_N$ while for $\pi\to \infty$, $e^2_{EB}(N) \to 1/\Delta^{2}_N$ while  $e^2_{M}(N)$ diverges to $+\infty$. As predicted earlier on,  for large values of the prior variance, $e^2_{M}(N)$ is obviously larger than $e^2_{EB}(N) $.\\
 However for large values of $N$ and small values of $\pi$  the mean square error of the Empirical Bayes estimator can be much larger  than $e^2_{M}(N)\simeq  1/\Delta^{2}_N$ since for $N\to \infty$ the square norm $\Delta^{2}_N$ must diverge   (a necessary condition for ergodicity) and hence $1/\Delta^{2}_N$  converges to zero and must be smaller than any preassigned parameter value $\theta_0^2$. Incidentally this guarantees that  the Bayes estimator is asymptotically unbiased (and  its MSE tends to zero).}
In conclusion we may say that for small values of the prior variance $\pi$ and large enough sample size the simpler marginal estimator can be  a better estimator than the Empirical Bayes. This is also  evident from the results of experiments, see table \ref{tab:res-var-par01} and \ref{tab:res-var-par02}.
 \end{example}

 \section{Hyperparameter estimation}\label{HypEst}
 
The  question is how should we estimate the hyperparameters, in particular the a priori variance~$\Pi$. Can we recover it from a suitably long chunk of observations of the process $\yb(t)$?  To address this question it will be convenient to  
consider the time evolution of the natural Bayesian estimator of the parameters of  an ARX model assuming for simplicity that the a priori mean $\mu$ is equal to zero. To this end denote the a priori variance $\sigma^2\Pi$ by $P_0$ and  return to  the finite-data  pseudo-linear regression model \eqref{ARXSol} where the end point $N$ is substituted by a generic stage index  $t$. \\
The  Bayes estimates at time $t$ are given by
\begin{align}
 \hat\xb(t) &=  \left({\Phi}_t^{\top}\Phi _t+\Pi^{-1}\right)^{-1}{\Phi}_t^{\top} \yb^{t}, \label{KFx}\\
 \hat P(t)& =\sigma^2 \left({\Phi}_t^{\top}\Phi_t+\Pi^{-1}\right)^{-1}\,,  \label{KFP}\\
 \hat{\sigmab}^{2}(t) &= \Frac{1}{t} \|\yb^t -\Phi_t \hat {\xb}(t)\|^{2}, \label{KFsigma}
 \end{align}
note that the first two expressions require knowledge of both  $\sigma^2$ and of the normalized a priori  variance of $\xb$, $\Pi$.\\
The unknown parameter $P_0=\sigma^2\Pi$  plays the role of initial covariance data in a Riccati-type algorithm for updating $P(t)$ which can be derived from   the Bayesian formula    \eqref{KFP}. Inserting  this parameter estimate in \eqref{KFP} leads to a natural estimator of $P_0^{-1}$ based on $N$ data,  given by 
\begin{align} \label{EstPrior}
&\hat{P}_0^{-1}(N)=P(N)^{-1}- \Frac{1}{\hat\sigma^2(N)}\Phi_N^{\top}\Phi_N\,\q \text{or}\q \notag\\
& \hat{\Pi}^{-1}(N)=\hat\sigma^2(N) P(N)^{-1}- \Phi_N^{\top}\Phi_N
\end{align}
which at first sight looks like  a reasonable solution. However,  since $N$ is usually assumed  large enough to make   $P(N)$ well defined and  invertible, a   warning is in order that  for large $N$ this difference could be quite small leading to errors amplification in the inverse.\\
 
\begin{rem} {\em Note also that    a sample of the joint process  $\{\yb(t),\ub(t)\}$, once inserted in the $p$-dimensional data vector \eqref{phi}, should satisfy the following limit relation:
\beq \label{Erg} 
\lim_{t \to \infty} \Frac{1}{t}\sum_{s=1}^t \varphib(s) \varphib(s)^{\top}= \Sigma
\eeq
almost surely,  where $\Sigma$ is $p\times p$ positive definite.  This condition is essentially   {\em second order ergodicity} of the joint process $\{\yb(t),\ub(t)\}$ which is always needed in time-series statistics and can be shown to hold under stability of the AR part of the model \cite[p. 510]{LPBook}.  It  implies that $\Phi_t^{\top}\Phi_t$  must also become positive definite for a suitable large $t$ and, in fact,  that this matrix must diverge as $t\to \infty$ at a linear rate with $t$ which in turn, by a limit argument in the Riccati recursion  (see  the appendix),  implies that 
\beq\label{ZeroP}
\lim_{t\to \infty} P(t) =0
\eeq
{\it independent of the initial condition} $P_0$. This even if  $\hat\sigmab_N^2(t)$   may converge to a constant (nonzero) positive value.}
\end{rem}
Hence, in  the limit of an {\em infinitely long} sequence of observations, the recover of $P_0$ would be ill-posed or  impossible. For  the limit \eqref{ZeroP} holds for any {\em positive semidefinite initial condition} $P_0$  and we may say that in the limit any positive semidefinite $P_0$ including $P_0=0$, could be a valid a priori variance. Therefore here we  incur in  {\em non-identifiability} of $P_0$. This  also agrees with the fact that when $\Phi_N^{\top}\Phi_N$ becomes very large the influence of the prior in the expressions \eqref{KFP} and \eqref{KFx}  becomes negligible  and the estimates tend to coincide with maximum likelihood  (i.e. plain least squares), where there is no prior involved.

To estimate  the unknown parameters $\theta,\sigma^2$ and $\Pi$ we shall need to work with finite data and use a different idea. We shall propose a sequential procedure  which  is essentially  a {\em Backward Conditional } Kalman Filter   adapted to the  ARX model.    \\
  For simplicity of notations  we shall  here assume that   the model \eqref{Randeff} is a purely autoregressive (AR) representation (no input). The generalization to full ARX models is straightforward as it only  requires more complicated notations and longer formulas. For ease of notation   it will be left to the reader, see however Corollary \ref{Cor} in the appendix.. 

The first step is to rewrite the pseudo-linear regression model  \eqref{ARX2} as a backward recursion. The representation of a process $\yb$ by such {\em backward models} has been discussed at length in the literature starting with the papers \cite{Lindquist-P-79},\cite{LP85} and used in various contexts. The basic step   is  spectral factorization and   the selection of the backward (i.e. antistable) spectral factor, as explained in the proof  of the following lemma.

\begin{lem}\label{Backwmodel}
Let the process $\{\yb(t)\}$ be represented by an $n$-th order  AR model with vector coefficient $\theta= \bmat a_1&\ldots&a_n\emat^{\top}$ and let
\beq \label{Barfi}
\bar\varphi(t): \bmat \yb(t+1) &\ldots &\yb(t+n)\emat^{\top},\qq t\in \,[0, N]
\eeq
then $\yb$ can be also represented by a backward recursion of the form
\beq \label{BackwAR}
\yb(t)= \bar\varphi(t)^{\top} \theta +\bar\wb(t), \qq t\in \Zbb
\eeq
where $\bar\wb(t)$ is   a stationary white noise of  finite  variance $\bar \sigma^2$, uncorrelated with all future variables $\{\yb(s);\, s> t \}$.
\end{lem}

\begin{pf}
Consider the  representation  of the  ARX model of order $n$ \eqref{ARX1} for $u\equiv 0$: 
\beq \label{AR}
\yb(t)= \sum_{k=1}^n a_k \yb(t-k) +\wb(t)
\eeq
where $\wb(t)$ is a white noise of variance $\sigma^{2}$, uncorrelated with all past variables $\{\yb(s)\mid s<t\}$, and   the polynomial $a(z^{-1})= 1-a_1z^{-1} -\ldots - a_n z^{-n}$ is such that     $z^{n}a(z)$ has all its zeros inside the unit circle.  The double-sided $z$-transform (denoted by hatted symbols) of the stationary zero-mean process $\yb$ described by \eqref {AR} can be written
$$
a(z^{-1})\hat y(z)= \hat w(z)
$$
so that the spectral density $S(z)= \E\hat y(z) \hat y(z^{-1})$  admits  the spectral factorization 
\beq
S(z)= \sigma^2\Frac{1}{a(z^{-1})} \Frac{1}{a(z)} \, \qquad z=e^{j\omega}\,.
\eeq
Since  the  polynomial   $z^{n}a(z)$ has all its zeros inside the unit circle, $a(z)$ has zeros of modulus greater than one, exactly in the reciprocal locations of those of $z^{n}a(z)$. While the transfer function $ \Frac{1}{a(z^{-1})}$ leads to a familiar {\em causal}  AR representation \eqref{AR}, the (antistable) spectral factor   $\Frac{1}{a(z)}$ leads to the  {\em anticausal AR representation}:
\beq \label{BackAR}
\yb(t)= \sum_{k=1}^n a_k \yb(t+k) +\bar{\wb}(t)
\eeq
where $\bar{\wb}(t)$ is another  white noise of   variance $\bar\sigma^{2}$ which is now uncorrelated with all {\em future } variables $\{\yb(s)\mid s>t\}$. The two white processes must be related by a transfer function as
\beq \label{ForwBackNoise}
 \bar{\wb}(t)= \Frac{a(z)}{a(z^{-1})} \wb(t)
\eeq
so that $\bar{\wb}(t)$ has also a positive constant spectral density. Once properly normalized, $\Frac{a(z)}{a(z^{-1})} $   is in fact  a conjugate rational {\em inner function} see e.g. \cite{LPBook}.\hfill$\Box$
\end{pf}

The recursion \eqref{BackAR} is run backwards in time starting at some  end point $t=N$, assuming that we have available a suitably long sequence of future measurements. Note that the backward AR model \eqref{BackAR} has the same deterministic parameters $\theta$ as the forward AR model \eqref{AR}. They both have a stochastic random effect version involving, in place of $\theta$,   the same  random parameter vector $\xb$, which, because of the relation \eqref{ForwBackNoise} is seen to be independent of the backward noise process $\bar\wb$. \\
In our problem  $\bar\varphib(t) $ is a function of future  measurements from time $t+1$ on  and here, to estimate the initial conditions  we shall  apply a {\em backward version} of  the conditional  generalized Kalman filter, which is described in the theorem below. 
\begin{thm}\label{BKF}
 Let $\bar{\yb}^{t}$ denote the vector made with the   available future components of $\yb$ at time $t$ ordered in  decreasing time order and define the matrix of future data at time $t$:
  \beq
 \bar{\Phi}_t:= \bmat \bar\varphib(N)^{\top}\\ \cdots \\\bar\varphib(t)^{\top}\emat
 \eeq
 which yields the backward pseudo-linear representation
 \beq 
 \bar{\yb}^{t}=  \bar{\Phi}_t\xb + \bar\wb^t\,.
 \eeq
 The conditional mean, $\bar{\xb}(t)$, and the normalized conditional variance of the random parameter vector~$\xb$ {\bf given the future data from time} $t$:
 \begin{equation}\label{BackBayesRec}
\bar{\xb}(t):= \E [\xb\mid \bar\yb^t]\,,\qquad  \bar P(t):=\frac{1}{\bar{\sigma}^2} \Var\,[ \,\xb\mid \bar\yb^t]
\end{equation}
satisfy the backward  recursion
\begin{equation}
\!\bar{\xb}(t\!-\!1) =\bar{\xb}(t)\! +\bar k(t\!-\!1) [\, \yb(t\!-\!1)-\bar\varphib(t\!-\!1)^{\top} \bar{\xb}(t)\,] \label{BKF1} 
\end{equation}
 where the {\em backward  gain vector} $\bar k(t\!-\!1)$ is given by
\begin{equation}
\bar k(t\!-\!1) = \bar P(t)\bar\varphib(t\!-\!1)\! \left [\bar\varphib(t\!-\!1)^{\top} \! \bar P(t) \bar\varphib(t\!-\!1) +1\right]^{-1} \label{BKFgain}
\end{equation}
while the backward covariance matrix   $\bar P(t)$ satisfies 
\begin{align}
\!\bar P(t-1)& \!= \!\bar P(t)-\!\bar P(t)\bar \varphib(t\!-\!1)\! \left [\bar \varphib(t\!-\!1)^{\top}\bar P(t) \bar \varphib(t\!-\!1) + 1\right]^{-1} \notag \\
& \bar \varphib(t\!-\!1)^{\top}\!\bar P(t)\,. \label{BRiccati}
\end{align}
The recursion
\beq \label{lambda}
\bar \lambda^2(t\!-\!1)= \Frac{t}{t\!+\!1}\bar \lambda^2(t) +\Frac{1}{t\!+\!1} [\yb(t\!-\!1)- \bar\varphib(t\!-\!1)^{\top}\bar {\xb}(t\!-\!1)\,]^2
\eeq
describes the  estimate of the conditional variance of the {\em backward innovation process} $\bar {\eb}(t):= \yb(t)-\bar \varphib(t)^{\top} \bar \xb(t)$. The corresponding estimate of $\sigma^2$ is expressed by the formula
\beq
\bar  {\sigmab}^2(t)= \Frac{\bar\lambda^2(t)}{\bar \varphib(t)^{\top}  \bar P(t) \bar \varphib(t) + 1}\,.
\eeq
This algorithm computes recursively in the backward direction,  the conditional mean, the normalized {\em backward} conditional variance $\bar P(t)$ of $\xb(t)$ given the future history of $\yb$ and the conditional variance $\bar{\sigmab}^2(t)$ of the   backward innovation. 
\end{thm}
The derivation of the algorithm is  reported in the appendix.  For completeness we recall that the  estimates \eqref{KFx}, \eqref{KFP}, \eqref{KFsigma} can  be updated recursively by  a {\bf forward }  conditional  Kalman filter algorithm,   adapted to Bayesian parameter estimation of ARX models. The idea  of this algorithm was already in \cite{Mayne-63} and \cite{Ho-Lee-64} but a rigorous justification   due to R. Liptser   is exposed in the book \cite[Chap 12]{Liptser-S-77}. A related but more general version  is presented in the paper \cite{Chen-K-S-89}.
\begin{rem} {\em One may object that  the terminal conditions $\bar x(N)$ and $\bar P(N)$ are not known.   However, under the structural conditions guaranteed to the backward model generating the data (in this case backward asymptotic stability), the backward  Kalman filter tends to forget the   terminal conditions and  its asymptotic behavior (in the backward direction)  turns out to be   independent of them for  a reasonable data length. \\
In extreme cases one could    couple a pair of  forward-backward Kalman filtering algorithms and first run a forward filter to get preliminary estimates of $\xb_N$ and $ P_N$ to be used as   starting data for the backward filter. They could be estimated by a  {\em forward} conditional  Kalman filter  whose asymptotic behaviour (now forward in time)  does not depends  on the knowledge of proper initial conditions $\xb_0$ and $P_0$. However  this procedure is seldom needed. }
\end{rem}

\section{A Comparison Example}\label{sec:compare_example}
We shall simulate  the simple  autoregressive  model
\beq \label{NominalAR}
\yb(t)= a_1 \yb(t-1)+a_2  \yb(t-2) +\wb(t)
\eeq
where $a_1=1.5,\, a_2=-0.7$ and $\wb$ is Gaussian i.i.d. with unit variance which generates an ergodic  stationary process. We run a forward conditional Kalman Filter started at various initial conditions $P_0$, stopping the simulation at $N_1=50, N_2=100, N_3=200$. Next
we compute    estimates of   the hyperparameters of our ARX model by the backward  procedure  using the the terminal conditions just described, the issue we are after  is to compare the performance of marginal Bayes with Empirical Bayes estimators for the specific model class under study. 

With these estimates we compute the scalar variances and MSE of the marginal and compare them with the Empirical Bayes MSE which can be computed using the differences of the estimates $\hat{\xb}(N)- \theta_0$ where $\theta_0$ is the true parameter+ the estimated Variance. We shall use the formula
\beq\label{MSEBayes}
\text{Emp. Bayes MSE}(N)= \|\E\hat{\xb}(N)- \theta_0\|^2 + \hat{\sigmab}^2(N) \tr P(N)
\eeq
 (implicitly using the conditional expected value given the past data instead of the plain expectation) and compare this with the MSE of the marginal estimate, equal to the  variance \eqref{varTheta} which, after substituting the estimate of the prior namely $\hat{P}_0(N)$ in place of $P_0$ becomes 
 \beq
\text{Marginal MSE}(N) = \hat{\sigmab}^2(N) \tr \{ [\Phi_N^{\top}\Phi_N]^{-1}+\tr\hat{P}_0(N)\}. 
\eeq
The results are reported in Table \ref{tab:res-fix-par}.
\begin{table}[h]
	\resizebox{\columnwidth}{!}{%
	\begin{tabular}{|c|c|c|c|c|c|c|}
		\hline
		\textbf{Initial $\boldsymbol{P_0}$} & \textbf{$\boldsymbol{\hat P_0}(N)$}  & \textbf{N}  & \makecell{\textbf{Marg. Bayes} \\ \textbf{Estimates}\\ $(\boldsymbol{\hat \theta_{M}}(N))$} & \makecell{\textbf{Marg. Bayes}\\ \textbf{MSE}} & \makecell{\textbf{Emp. Bayes}\\ \textbf{Estimates}\\ $(\boldsymbol{\hat x_{B}}(N))$ }  & \makecell{\textbf{Emp. Bayes} \\\textbf{MSE}} \\ \hline \hline
		$\begin{bmatrix}.01 & 0\\ 0 & .01\end{bmatrix}$   & $\begin{bmatrix} .01201 & -.00008\\ -.00008 & .01194\end{bmatrix}$  & 50  & $\begin{bmatrix}1.45760 \\ -.80157\end{bmatrix}$   & 0.03674      & $\begin{bmatrix} .72873\\-.16379 \end{bmatrix}$			  & 0.90160   \\ \hline
		$\begin{bmatrix}.01 & 0\\ 0 & .01\end{bmatrix}$   & $\begin{bmatrix} .01442 & -.00009\\ -.00009 & .01434\end{bmatrix}$  & 100 & $\begin{bmatrix}1.49922 \\ -.73953\end{bmatrix}$               & 0.04079             & $\begin{bmatrix} .98814\\-.26275 \end{bmatrix}$ & 0.46583   \\ \hline
		$\begin{bmatrix}.01 & 0\\ 0 & .01\end{bmatrix}$   & $\begin{bmatrix} .01417 & -.00009\\ -.00009 & .01409\end{bmatrix}$  & 200 & $\begin{bmatrix}1.48003 \\ -.72652\end{bmatrix}$               & 0.03201             & $\begin{bmatrix} 1.12962\\-.40115 \end{bmatrix}$ & 0.23242   \\ \hline
		$\begin{bmatrix}.08 & 0\\ 0 & .08\end{bmatrix}$     & $\begin{bmatrix} .63853 & -.03381\\ -.03381 & .60759\end{bmatrix}$ & 50  & $\begin{bmatrix}1.45760 \\ -.80157\end{bmatrix}$               & 1.12862             & $\begin{bmatrix} 1.19486\\-.55851 \end{bmatrix}$ & 0.15418   \\ \hline
		$\begin{bmatrix}.08 & 0\\ 0 & .08\end{bmatrix}$     & $\begin{bmatrix} .60511 & -.03204\\ -.03204 & .57579\end{bmatrix}$ & 100 & $\begin{bmatrix}1.49922 \\ -.73953\end{bmatrix}$               & 1.22632             & $\begin{bmatrix} 1.40620\\-.64994 \end{bmatrix}$ & 0.07475   \\ \hline
		$\begin{bmatrix}.08 & 0\\ 0 & .08\end{bmatrix}$     & $\begin{bmatrix} .47637 & -.02522\\ -.02522 & .45329\end{bmatrix}$ & 200 & $\begin{bmatrix}1.48003 \\ -.72652\end{bmatrix}$               & 1.00381             & $\begin{bmatrix} 1.42646\\-.67597 \end{bmatrix}$ & 0.03051   \\ \hline 
	\end{tabular}%
	}
	\caption{Results of simulations with fixed parameters.}
	\label{tab:res-fix-par}
\end{table}

In the table we have reported  simulation results  corresponding to  two initial conditions (a priori variance matrices) $P_{0,i}=\sigma_i^2\Pi_i$ (with $\sigma_i^2=1$): 
$$  
P_{0,1}=\begin{bmatrix}0.01 & 0\\ 0 & 0.01\end{bmatrix},\quad  
P_{0,2}=\begin{bmatrix}0.08 & 0\\ 0 & 0.08\end{bmatrix},  
$$ 
which, once compared with the relative estimates, can be classified as "medium" and "large" size, respectively. For instance, in the case of the medium one with $N=200$ samples the estimated prior variance gives a standard deviation $\sigma\approx 0.1187$ for both parameters $[a_1,a_2]$  and hence all the values of the two parameters are in the range $a_1\pm3\sigma$ and $a_2\pm3\sigma$, respectively, that  seems to be a not so large deviation. Applying the same reasoning for the case of large prior then the standard deviation becomes $\sigma\approx 0.69$ for $a_1$ and $\sigma\approx 0.673$ for $a_2$ which  seem much higher. 
For each $P_{0,i}$ we have calculated the relative estimate $\hat P_0$ using the backward Kalman  of Theorem \ref{BKF}. 
In  the right columns  are listed the estimates of the parameters and the MSE for both  the Marginal Bayes and Empirical Bayes  procedure. As we can see, for a medium-size true prior, the Marginal estimates are weakly dependent on the number of samples $(N)$ and almost consistent, indeed with a small bias. Conversely, the Bayes estimates are strongly dependent on the number of samples $N$, the higher $N$ the better the estimates. For large $P_0$ we observe the same behavior of the parameter estimates, but with a better quality of the Bayes ones. \\
It is quite evident that  for medium-size prior variance matrix  the Marginal Bayes MSE is definitely smaller than for the Empirical Bayes  and consequently Marginall Bayes  behaves  better, as it is also evident looking at the parameter estimates. Observe the decrease of the MSEs with the increase of the number of samples ($N$) as expected from the theory. \\
Finally, for a considerably  larger initial prior, one can clearly observe that  the MSE of the Empirical  Bayes   estimate becomes  smaller than that of the Marginal Bayes estimates. Also  the prior variance estimates are very clearly biased.

\begin{figure}[h!]
	\centering
	\begin{subfigure}[b]{.9\linewidth}
		\includegraphics[width=\linewidth]{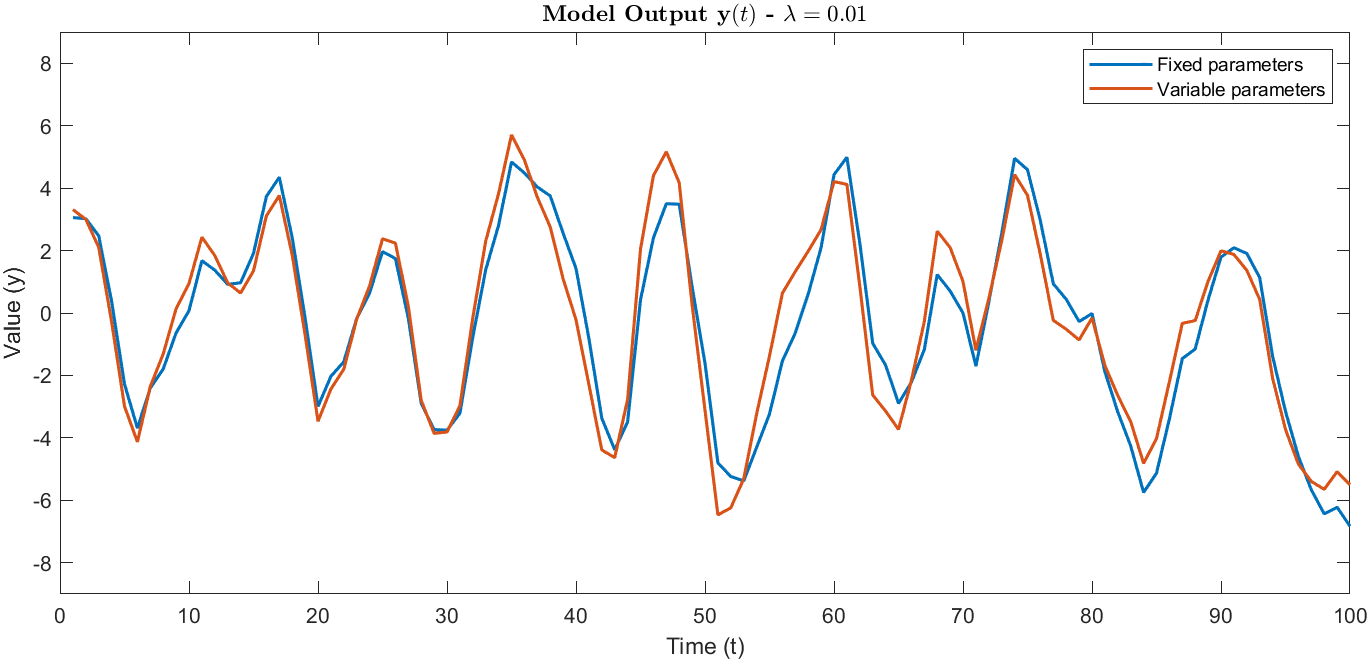}
	\end{subfigure}
	\begin{subfigure}[b]{.9\linewidth}
		\includegraphics[width=\linewidth]{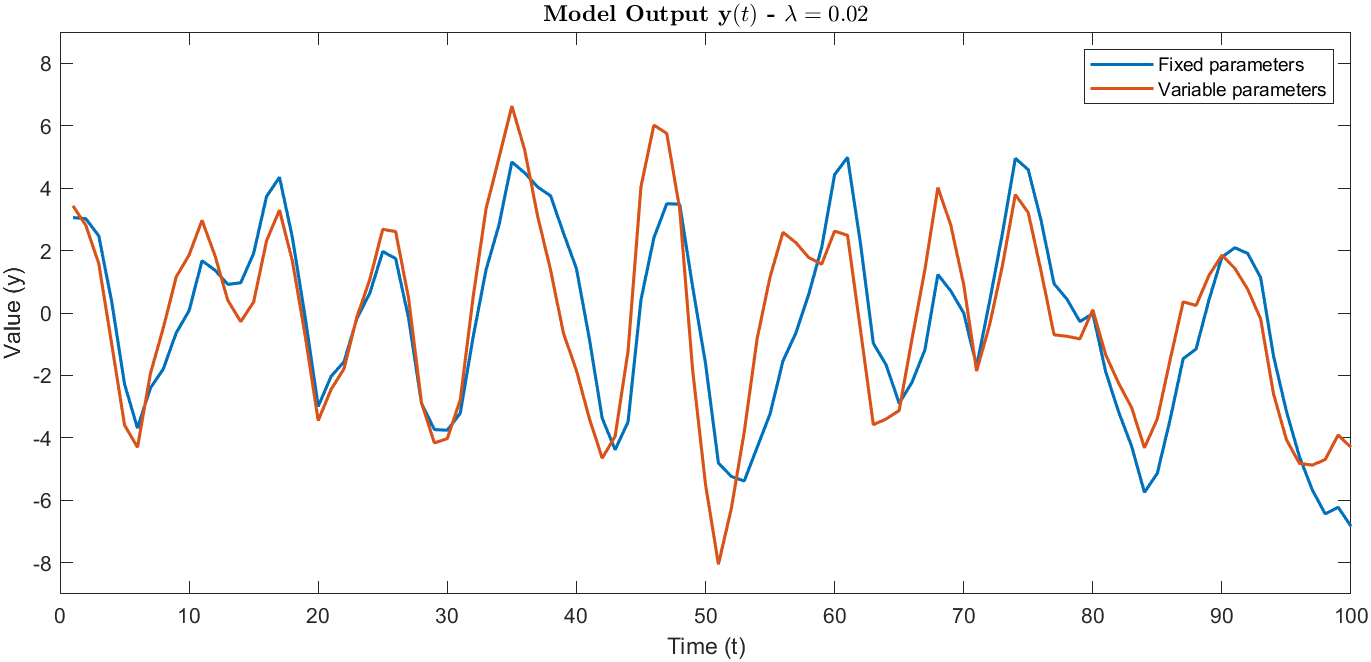}
	\end{subfigure}
	\caption{Outputs corresponding to N=100 and $\lambda=0.01, 0.02$ for variable parameters. The plots with fixed parameters in the two graphics are identical.}
	\label{fig:output-01}
\end{figure}

Next, we  consider  a more realistic example  generating the data $\yb(t)$  by simulating the model \eqref{NominalAR} assuming  {\em  slowly varying random parameters}.  We describe the parameter variation by the following  linear  random model:
$$
\bmat a_1(t+1)\\a_2(t+1)\emat= \bmat  1.5\\ -.7\emat+  \bmat .98 &0\\0&.97\emat \bmat a_1(t)-1.5\\a_2(t)+0.7\emat + \lambda \bmat \wb_1(t)\\ \wb_2(t)\emat
$$
where $\wb_1$ and $\wb_2$ are independent white noises with a small  standard deviation say $\lambda$ equal to $0.01$ or $0.02$. The simulated  behaviors of the output, compared with the   nominal system, are shown in Figure \ref{fig:output-01}, where the data correspond to  the same trajectory of the generating noise $\wb$.

By repeating  the calculations of the previous example, we obtain the results shown in the tables \ref{tab:res-var-par01} and \ref{tab:res-var-par02}  which correspond to two different standard deviations $\lambda$ of the parameter fluctuation noise.
In this case the behavior of the MSEs and estimates are similar to those with fixed parameters, instead we can observe the increases of the bias in the prior variance estimates, i.e. $\hat P_0(N)$.

\begin{table}[]
	\resizebox{\columnwidth}{!}{%
		\begin{tabular}{|c|c|c|c|c|c|c|}
			\hline
			\textbf{Initial $\boldsymbol{P_0}$} & \textbf{$\boldsymbol{\hat P_0}(N)$}  & \textbf{N}  & \makecell{\textbf{Marg. Bayes} \\ \textbf{Estimates}\\ $(\boldsymbol{\hat \theta_{M}}(N))$} & \makecell{\textbf{Marg.} \\\textbf{Bayes}\\ \textbf{MSE}} & \makecell{\textbf{Emp. Bayes}\\ \textbf{Estimates}\\ $(\boldsymbol{\hat x_{B}}(N))$ }  & \makecell{\textbf{Emp. Bayes} \\\textbf{MSE}} \\ \hline \hline
			
			$\begin{bmatrix}.01 & 0\\ 0 & .01\end{bmatrix}$   & $\begin{bmatrix} .01130 & -.00008\\ -.00008 & .01110\end{bmatrix}$ & 50         & $\begin{bmatrix} 1.47650\\-.84354 \end{bmatrix}$                        & 0.03319                   & $\begin{bmatrix} .78642\\-.23015 \end{bmatrix}$        & 0.74723            \\ \hline
			
			$\begin{bmatrix}.01 & 0\\ 0 & .01\end{bmatrix}$   & $\begin{bmatrix} .01508 & -.00008\\ -.00008 & .01484\end{bmatrix}$ & 100        & $\begin{bmatrix} 1.48029\\-.76864 \end{bmatrix}$                        & 0.04286                   & $\begin{bmatrix} 1.01273\\-.33566 \end{bmatrix}$        & 0.38233            \\ \hline
			
			$\begin{bmatrix}.01 & 0\\ 0 & .01\end{bmatrix}$   & $\begin{bmatrix} .01370 & -.00013\\ -.00013 & .01356\end{bmatrix}$ & 200        & $\begin{bmatrix} 1.46703\\-.73076 \end{bmatrix}$                        & 0.03106                   & $\begin{bmatrix} 1.12586\\-.41472 \end{bmatrix}$        & 0.22698            \\ \hline
			
			$\begin{bmatrix}.08 & 0\\ 0 & .08\end{bmatrix}$     & $\begin{bmatrix} .48553 & -.02840\\ -.02840 & .41492\end{bmatrix}$ & 50         & $\begin{bmatrix} 1.47650\\-.84354 \end{bmatrix}$                        & 0.82156                   & $\begin{bmatrix} 1.33350\\-.70849 \end{bmatrix}$        & 0.09042            \\ \hline
			
			$\begin{bmatrix}.08 & 0\\ 0 & .08\end{bmatrix}$     & $\begin{bmatrix} .63864 & -.02725\\ -.02725 & .55624\end{bmatrix}$ & 100        & $\begin{bmatrix} 1.48029\\-.76864 \end{bmatrix}$                        & 1.37825                   & $\begin{bmatrix} 1.40196\\-.69341 \end{bmatrix}$        & 0.06532            \\ \hline
			
			$\begin{bmatrix}.08 & 0\\ 0 & .08\end{bmatrix}$     & $\begin{bmatrix} .45993 & -.03710\\ -.03710 & .42031\end{bmatrix}$ & 200        & $\begin{bmatrix}1.46703\\ -.73076\end{bmatrix}$                         & 0.85930                   & $\begin{bmatrix} 1.41844\\-.68503 \end{bmatrix}$         & 0.02890            \\ \hline
		\end{tabular}%
	}
	\caption{Results of simulations with variable parameters and $\lambda=0.01$.}
	\label{tab:res-var-par01}
\end{table}

\begin{table}[h!]
	\resizebox{\columnwidth}{!}{%
		\begin{tabular}{|c|c|c|c|c|c|c|}
			\hline
			\textbf{Initial $\boldsymbol{P_0}$} & \textbf{$\boldsymbol{\hat P_0}(N)$}  & \textbf{N}  & \makecell{\textbf{Marg. Bayes} \\ \textbf{Estimates}\\ $(\boldsymbol{\hat \theta_{M}}(N))$} & \makecell{\textbf{Marg.} \\\textbf{Bayes}\\ \textbf{MSE}} & \makecell{\textbf{Emp. Bayes}\\ \textbf{Estimates}\\ $(\boldsymbol{\hat x_{B}}(N))$ }  & \makecell{\textbf{Emp. Bayes} \\\textbf{MSE}} \\ \hline \hline
			
			$\begin{bmatrix}.01 & 0\\ 0 & .01\end{bmatrix}$   & $\begin{bmatrix} .01126 & .00007\\ .00007 & .01077\end{bmatrix}$ & 50         & $\begin{bmatrix} 1.49695\\-.87744 \end{bmatrix}$                        & 0.03234                   & $\begin{bmatrix} .83624\\-.28188 \end{bmatrix}$        & 0.63180            \\ \hline
			
			$\begin{bmatrix}.01 & 0\\ 0 & .01\end{bmatrix}$   & $\begin{bmatrix} .01463 & -.00003\\ -.00003 & .01426\end{bmatrix}$ & 100        & $\begin{bmatrix} 1.45567\\-.79186 \end{bmatrix}$                        & 0.04154                   & $\begin{bmatrix} 1.02877\\-.39871 \end{bmatrix}$        & 0.32363            \\ \hline
			
			$\begin{bmatrix}.01 & 0\\ 0 & .01\end{bmatrix}$   & $\begin{bmatrix} .01625 & -.00008\\ -.00008 & .01550\end{bmatrix}$ & 200        & $\begin{bmatrix} 1.47271\\-.73062 \end{bmatrix}$                        & 0.03779                   & $\begin{bmatrix} 1.15014\\-.42892 \end{bmatrix}$         & 0.20237            \\ \hline
			
			$\begin{bmatrix}.08 & 0\\ 0 & .08\end{bmatrix}$     & $\begin{bmatrix} .53880 & .02513\\ .02513 & .34930\end{bmatrix}$   & 50         & $\begin{bmatrix} 1.49695\\-.87744 \end{bmatrix}$                        & 0.83556                   & $\begin{bmatrix} 1.37114\\-.75324 \end{bmatrix}$        & 0.06840            \\ \hline
			
			$\begin{bmatrix}.08 & 0\\ 0 & .08\end{bmatrix}$     & $\begin{bmatrix} .58419 & -.01039\\ -.01039 & .46649\end{bmatrix}$ & 100        & $\begin{bmatrix}1.45567\\ -.79186\end{bmatrix}$                         & 1.24102                   & $\begin{bmatrix} 1.38857\\-.72718 \end{bmatrix}$         & 0.05562            \\ \hline
			
			$\begin{bmatrix}.08 & 0\\ 0 & .08\end{bmatrix}$     & $\begin{bmatrix} .63437 & -.02433\\ -.02433 & .39928\end{bmatrix}$ & 200        & $\begin{bmatrix}1.47271\\ -.73062\end{bmatrix}$                         & 1.08385                   & $\begin{bmatrix} 1.43353\\-.69226 \end{bmatrix}$         & 0.02675            \\ \hline
		\end{tabular}%
	}
	\caption{Results of simulations with variable parameters, $\lambda=0.02$.}
	\label{tab:res-var-par02}
\end{table}
The wording "medium" and a "large" a priori variance in the simulations  is based on a comparison obtained referring to  the   range of the estimated empirical standard deviations about the mean of the parameters. When these two classes  of   prior variance are suitably fixed   one can obtain  a fair comparison between the two techniques studied above. Once a reasonable ``medium and large" prior variance is assigned as initial matrices,  one can clearly see  that the Empirical Bayes gives a nice estimate of parameters along a small MSE with respect to the Bayes a posteriori.\\
 In the case of randomly varying parameters one could however observe some strange fluctuations due to  non-linearity in the data. Nonetheless, as the number of samples increase the MSE decreases, in fact, as also the theory suggests. \\
For a relatively large prior variance the Marginal Bayes is no longer a good alternative. The MSE becomes larger than that of the Empirical Bayes. Thus, with a large amount of data one should instead  choose    the Empirical Bayes. 


\section{Conclusions}
In this paper we  describe and compare the performance of the Marginal Bayes with that of  the Empirical  Bayes estimators for  ARX models of stationary time-series. We also propose a method to estimate the hyperparameters for both  estimators,   a non-trivial problem which may possibly turn out to be ill-posed. The results of the  simulations  of Section \ref{sec:compare_example}, both with fixed model  parameters and with randomly varying parameters, seem to confirm the  theoretical comparison of the two estimation principles  discussed in Section \ref{sec:emp_bayes_estim}. A rough indication being that the simpler  Marginal Bayes method can perform  better for small sample size and small a priori variance, which is somewhat reasonable. We remark that all the results here are {\em non-asymptotic} and are valid for   finite sample size which may often be the realistic situation in practice.

\bibliographystyle{plain}
\bibliography{biblio-BayesARX}
\appendix
\section{The Backward  Kalman filter for Conditionally Gaussian ARX models}\label{CondGaussKF}
In this appendix, for ease of reference to the reader,  we shall present in a self-contained way the  backward version of the  {\em Conditionally Gaussian Kalman filter} which is a well known algorithm adapted to Bayesian parameter estimation of ARX models, a backward version of that    in \cite{Mayne-63}\cite{Ho-Lee-64} and \cite[Chap 12]{Liptser-S-77}. 

We shall use without further comments the notation introduced in Lemma \ref{Backwmodel} and in the following Theorem \ref{BKF}. Consider the ARX model \eqref{ARX1}   written as a  backward pseudo-linear regression (33)
\begin{equation}\label{ARX2bis}
 \yb(t) =    \bar{\varphib}(t)^{\top}\, \theta   + \bar{\wb}(t)\,,\qquad t\in \Zbb 
\end{equation}
where
 $$
 \bar{\varphib}(t)^{\top}=\bmat \yb (t+1) &\ldots & \yb(t+n)\emat^{\top} ,\qquad t\in \Zbb \,.
 $$
We   assign  a Gaussian a priori distribution to the parameter which becomes a random $p$-dimensional vector denoted $\bar{\xb}$ 
$$
\bar{\xb}\sim {\script N}(\bar\mu ,\bar{\Sigma})
$$
with $\bar\mu $ some ``nominal'' mean value and $\bar{\Sigma} $ a variance matrix which is usually unknown.  The noise $\bar{\wb}(t)$ is also   Gaussian i.i.d. with variance $\bar{\sigma}^2$, independent of  $\bar{\xb}$ for all $t$.

For simplicity we shall initially assume that $\bar{\varphib}(t)$ is only a function of the future $\bar{\yb}^{t}$ so that the original model is    purely Auto Regressive\footnote{In fact the derivation which follows is valid for $ \bar{\varphib}(t)$ being an {\em arbitrary function of the future outputs}}. At the end we shall consider  generalizations of this model and   allow $ \bar{\varphib}(t)$ to depend also on future inputs $\ub$. In this case we shall need to require that $\ub(t)$ and $\wb(s)$ are independent for all $t,s \in \Zbb$.

     The model \eqref{ARX2bis} is a backward (antistable) difference equation which can be solved recursively starting from some  string oerivatiof $n$  final values at an arbitrary final time,   which by stationarity we can and shall assume   equal to  zero   so that   the backward recursion \eqref{ARX2bis} once solved starting from the final data $\bar{\varphib}(0)$ yelds a solution
 $$
\yb(t)= h(\bar{\varphib}(0)\,\bar{\xb},\,\bar{\wb}^t)\,; \qquad t\leq 0
$$
  which is a function of the parameter, the final conditions (at time zero),  and the past noise from time zero down to time $t<0$. From the independence of $\bar{\wb}(t\!-\!1)$ and $\bar{\wb}^t$ we see immediately that,
\begin{prop}\label{lem:uncorr}
For all $t\leq 0$ the random variable  $\bar{\wb}(t\!-\!1)$ is independent of the future  observations $\bar{\yb}^t$; in fact it is also independent of $(\,\bar{\yb}^t,\,\bar\xb\,)$.
\end{prop}
Recall  that a random variable $ \xib$ is {\em conditionally Gaussian} given a family of random variables $\{\zb_{\alpha}\,;\, \alpha\in A\}$ if $\xib$ admits a conditional distribution given $\{\zb_{\alpha}\,;\, \alpha\in A\}$ which is  Gaussian. Naturally the  mean and variance of this distribution will be the conditional mean and variance of $\xib$ given $\{\zb_{\alpha}\,;\, \alpha\in A\}$.

 By stacking the system equations \eqref{ARX2bis} ordered for decreasing time $t=0,-1,-2,\ldots$ we obtain a relation among random vectors of the form
\begin{equation}\label{CondLin}
\bar{\yb}^{t}= \bmat \bar{\varphib}(-1)^{\top}\\ \ldots \\\bar{\varphib}(t)^{\top}\emat \bar{\xb}+ \bar{\wb}^t \,:=\,\bar{\Phib}_t\,\bar \xb+\bar{\wb}^t\,,\qq t<0
\end{equation}
where the matrix $\bar{\Phib}_t$ is a function of the final conditions and   past outputs from (negative) time  $t-1$ up to time $-1$.

\begin{thm}\label{thmCondGauss}
Assume $|t|$ is large enough so that $\bar\Phib_t$ has almost surely a left inverse $\bar\Phib_t^{-L}$. Then,  the random variables $(\,\yb(t\!-\!1),\,\bar\xb\,) $ are jointly  conditionally Gaussian given $\bar{\yb}^{t}$.
\end{thm}
\begin{pf}
We shall first show that $p( \bar\xb \mid \bar{\yb}^{t})$  is a conditionally Gaussian distribution. Left-multiply \eqref{CondLin} by $\bar\Phib_t^{-L}$ (which only depends on $\bar{\yb}^{t-1}$) to get 
$$
\bar\xb=  \bar\Phib_t^{-L} \bar\yb^t- \bar\Phib_t^{-L}\bar\wb^t
$$
which shows that the conditional distribution of $\bar\xb$  given $\bar\yb^t$ is actually Gaussian, 
\footnote{Since here  $\bar\xb$ depends on the choice of the left inverse, more correctly one should say: any random vector $\bar\xb$ satisfying \eqref{CondLin}.}  with (conditional) mean vector $ \bar\Phib_t^{-L}\yb^t$ and conditional variance   equal to \\$\bar\sigma^2  \bar\Phib_t^{-L}\,[\, \bar\Phib_t^{-L}\,]^{\top}$.\\
Then the statement   follows from Bayes rule\\
\begin{align*}
p( y(t\!-\!1),\,\bar x \mid \bar\yb^{t}=\bar y^{t})=&\, p(y(t\!-\!1) \mid \bar\xb=\bar x,\, \bar\yb^{t}=\bar y^{t}) \\ 
&\cdot p( \bar x \mid \bar\yb^{t}= \bar y^{t})
\end{align*}
since the first factor on the right is clearly a conditional Gaussian distribution with mean $\bar\varphib(t\!-\!1)^{\top}\theta$ and variance equal to $\var \{\bar \wb(t\!-\!1)\}$. If we condition with respect to some fixed observation string $\bar\yb^t(\omega)= \bar y^t$ these are  just   usual regular Gaussian densities. 
\end{pf}

In spite of its appearance, the model \eqref{CondLin} is non-linear and it is not immediately clear what could be a reasonable estimation strategy.  Consider then the following  
\begin{prob}
Find a recursive updating algorithm to compute the conditional mean and conditional variance of the random parameter vector $\bar\xb$:
\begin{equation}\label{BayesRec}
\hat{\bar\xb}(t):= \E [\bar\xb\mid \bar\yb^t]\,,\qquad \bar \Sigma(t):= \Var\,[ \,\bar\xb\mid \bar\yb^t]\,.
\end{equation}
\end{prob}
We shall   describe a {\em Bayesian recursive solution} of this problem which is essentially a backward version of the Conditional Kalman filter referred to in the beginning of this appendix.

 We shall rely on Theorem  \ref{thmCondGauss} and on the full rank assumption of $\bar\Phib_t$. Consider the conditional expectation of an arbitrary  random variable $\xib$ given two other random variables, $\yb_1, \yb_2$ the first of which is kept fixed while the second may vary. We shall need to consider the regression function $\E[\, \xib\mid \yb_1,\yb_2]$, which is by definition a measurable function of both variables.  When   $\yb_1$ is kept fixed, it can be  considered   as a function of $\yb_2$ only and  denote it by the symbol $ \E_{ \yb_1}[\,\xib\mid \yb_2]$. Obviously with this convention 
 $ 
 \E_{ \yb_1}[\,\xib\,]= \E [\,\xib\mid \yb_1]$. 
 Suppose $\xib$ is conditionally Gaussian given  $(\yb_1, \yb_2)$, then the iterated Gaussian  conditional expectation formula  yields
\begin{align}
\E[\, \xib\mid \yb_1,\yb_2] =&\, \E_{ \yb_1}[\,\xib\,] + \Cov \{\xib, \yb_2\mid \yb_1\} \Var \{  \yb_2\mid \yb_1\}^{-1} \nonumber\\
& \cdot \Cov \{\yb_2, \xib \mid \yb_1\} \,\{ \yb_2 - \E [\,\yb_2\mid \yb_1]\,\}  \label{TwostepCE}
\end{align}
which can be justified just by thinking that the conditional density with respect to both variables $p(x\mid y_1,y_2)$    is the  Gaussian  density $ p_{y_1}(x):= p(x\mid y_1)$ conditioned with respect to $\yb_2$. Consider now the estimate at time $t-1$
$$
\hat{\bar\xb}(t\!-\!1):= \E [\bar\xb\mid \yb(t\!-\!1),\,\bar\yb^t]=  \E_{ \bar\yb^t}[\,\bar\xb\mid \yb(t\!-\!1) \,]
$$
where the operator $\E_{\bar \yb^t}$ is as defined above.   Then applying formula  \eqref{TwostepCE}  we obtain
\begin{align}
\E [\bar\xb\mid \bar\yb(t\!-\!1),\,\bar\yb^t] =&\, \E_{ \bar\yb^t}[\, \bar\xb\,] + \nonumber\\
&+ \{\, \Cov \{\bar\xb,  \yb(t\!-\!1)\mid \bar\yb^t\} \nonumber\\ 
&\cdot \Var \{  \yb(t\!-\!1)\mid \bar\yb^t\}^{-1} \nonumber\\
&\cdot \Cov \{ \yb (t\!-\!1), \bar\xb\mid \bar\yb^t\} \nonumber\\
&\cdot \{ \yb(t\!-\!1) - \E [\, \yb(t\!-\!1)\mid \bar\yb^t]\,\}\,\}  \label{TwostepCE1}
\end{align}
where
$$
 \E_{ \bar\yb^t}[\,\bar\xb\,]= \hat{\bar\xb}(t)\,,\qquad \E [\,\yb(t\!-\!1)\mid \yb^t]= \bar\varphib(t\!-\!1)^{\top}\hat{\bar\xb}(t)
$$
the last equality following since $\bar\wb(t\!-\!1)$ and $\bar\yb^t$ are independent (Lemma \ref{lem:uncorr}). The last equation involves the (one step ahead) {\em backward-predictor} of $\yb(t\!-\!1)$ given $\bar\yb^t$ which is  denoted $\hat{\yb}(t-1\!\mid \!t)$. The estimator \eqref{TwostepCE1} is a linear function of the (one-step) {\bf backward prediction error}
\begin{align}
\bar\eb(t-1)&:= \yb(t-1)- \hat{\yb}(t-1\!\mid \!t) \nonumber\\ 
 &= \bar\varphib(t-1)^{\top}[\, \bar\xb-\hat{\bar\xb}(t)\,] +\bar\wb(t-1)
\end{align}
which is just the part of $\yb(t\!-\!1)$ which is unpredictable  based on the future $\yb^t$. By the orthogonality principle $\bar\eb$ is a process with conditionally uncorrelated (and hence independent) variables. It is called the {\bf backward (conditional)  innovation process} of $\{\yb(t)\}$. Its conditional variance is   
\begin{align*}
\var [\bar\eb(t\!-\!1)\! \mid \bar\yb^{t}] &= \var [\, \yb(t\!-\!1)\mid \bar\yb^{t} ] \\ 
 &= \E\{[\bar\varphib(t\!-\!1)^{\top}(\xb-\hat{\bar\xb}(t)) +\bar\wb(t\!-\!1)\,]^2\mid  \bar\yb^{t}\}\\
&=\bar\varphib(t\!-\!1)^{\top}\bar \Sigma(t) \bar\varphib(t\!-\!1) +\bar\sigma^2
\end{align*}
For the recursive estimation formula we need  the following covariance matrix   obtained  using the rule \eqref{TwostepCE}:
\begin{align}
&\Cov\{\bar\xb, \yb(t\!-\!1)\mid \bar\yb^t\} = \nonumber\\
&= \E\{[\, \bar\xb-\hat{\bar\xb}(t)\,] [\, \yb(t\!-\!1) - \hat{\yb}(t\!-\!1\!\mid\! t)\,] \mid \yb^t\}\nonumber\\
&= \E\{[\, \bar\xb-\hat{\bar\xb}(t)\,] [\, \bar\xb-\hat{\bar\xb}(t)\,] ^{\top}\bar\varphib(t\!-\!1)\mid \yb^t\}\nonumber\\ 
&\q\,+ \E\{[\, \bar\xb-\hat{\bar\xb}(t)\,] \bar\wb(t\!-\!1)\mid \bar\yb^t\}\nonumber\\
&=\E\{[\, \bar\xb-\hat{\bar\xb}(t)\,] [\, \bar\xb-\hat{\bar\xb}(t)\,] ^{\top} \mid \bar\yb^t\}\bar\varphib(t\!-\!1)\nonumber\\ 
&= \bar\Sigma(t)\bar\varphib(t\!-\!1) \,.
\end{align}
The last equality follows from the independence of $\bar\xb$ and $\bar\wb(t)$ and Lemma \ref{lem:uncorr}. Using the conditional variance formula $ \var [\, \yb(t\!-\!1)\mid \yb^{t} ]$ which has been  computed above,  the updating formula \eqref{TwostepCE1} can therefore be written as
\begin{equation}
\!\hat{\bar\xb}(t\!-\!1) =\hat{\bar\xb}(t)\! +\bar k(t\!-\!1) [\, \yb(t\!-\!1)-\bar\varphib(t\!-\!1)^{\top} \hat{\bar\xb}(t)\,] 
\end{equation}
 where the {\em backward gain vector} $\bar k(t\!-\!1)$ is given by
\begin{equation} \label{KFgain}
\bar k(t\!-\!1) = \bar\Sigma(t)\bar\varphib(t\!-\!1)\! \left [\bar\varphib(t\!-\!1)^{\top} \! \bar\Sigma(t) \bar\varphib(t\!-\!1) +\bar\sigma^2\right]^{-1} 
\end{equation}
The recursion is updated for decreasing times and  is driven by the backward innovation $\bar\eb(t\!-\!1)$.The initial condition is in principle equal to  $\hat{\bar\xb}(0)=\E\bar\xb$.\\
We still need an updating equation for $\bar\Sigma(t)$. Using again the  iterated conditioning formula \eqref{TwostepCE} and \eqref{TwostepCE1}  one gets
\begin{align*}
&\bar\Sigma(t\!-\!1)=\\ 
&=\Var [\,\bar\xb\mid \bar\yb^{t\!-\!1}]\\
&= \Var  [\,\bar\xb\mid \yb^t\,] -\{\Cov\{\xb, \yb(t\!-\!1)\mid \bar\yb^t\} \var [\, \yb(t\!-\!1)\mid \bar\yb^{t} ]^{-1} \\ 
&\q\cdot \Cov\{ \yb(t\!-\!1), \bar\xb\mid \bar \yb^t\}\,\}\\
&= \Sigma(t)- \{\, \!\Sigma(t)\varphib(t\!-\!1)\! \left[\varphib(t\!-\!1)^{\top} \! \bar\Sigma(t) \bar\varphib(t\!-\!1) +\sigma^2 \right]^{-1} \\
&\q\cdot \bar\varphib(t\!-\!1)^{\top}\!\bar\Sigma(t)\,\}
\end{align*} 
now with initial condition the a priori covariance $\bar\Sigma(0)=\Var [\,\bar \xb\,]$. 

 The reader can easily get a {\em normalized version} of this  conditional backward Kalman filter  algorithm   by {\em normalization} of the Covariance matrix dividing $\bar\Sigma(t)$ by $\bar\sigma^2$ thereby   obtaining exactly the formulas in Theorem \ref{BKF}.

Note that  the gain and the variance matrix are functions of the future data $\bar\yb^t$ making the algorithm a truly non-linear recursion. 

 There are various extensions of the algorithm to more complicated models. One easy step is to consider ARX models where the input process $\ub(t)$ enters as in \eqref{ARX1} and the parameter is now $n+m$-dimensional. Since  in general for physical reasons there cannot be instantaneous effect of the input $\ub(t)$ on the variable $\yb(t)$   the input parameter $b_0$ in the model is normally set to zero. Hence the future information available at time $t-1$ is   constituted by the joint input-output sequence
$\zb^t:= (\bar \ub^{t}, \bar\yb^{t})$.  The reader can work out the previous derivation considering all conditional expectations with respect to this joint information flow assuming that the input and the noise processes are independent.

\begin{cor}\label{Cor}
Consider the ARX model \eqref{ARX1} with a Gaussian noise  $\wb$ independent of the input process $\ub$. Then the estimator $\hat \xb(t)$ which minimizes the conditional error variance $\Sigma(t):= \Var  [\,\xb\mid \zb^t\,] $ evolves in time according to the same recursion, that is 
\begin{equation}
\!\hat{\xb}(t\!+\!1) =\hat{\xb}(t)\! +k(t\!+\!1) [\, \yb(t\!+\!1)-\varphib(t\!+\!1)^{\top} \hat{\xb}(t)\,] \label{KF1} 
\end{equation}
 where the {\em gain vector} $k(t\!+\!1)$ is given by
\begin{equation}
k(t\!+\!1) = \Sigma(t)\varphib(t\!+\!1)\! \left [\varphib(t\!+\!1)^{\top} \! \Sigma(t) \varphib(t\!+\!1) +\sigma^2\right]^{-1} \label{KF1gain}
\end{equation}
The process $\eb(t\!+\!1):= \yb(t\!+\!1)-\varphib(t\!+\!1)^{\top} \hat{\xb}(t)$ driving the recursion  is   the one step ahead prediction error of $\yb(t+1)$ given the past $\zb^t$ (that is the innovation). The initial condition can be taken as $\hat{\xb}(0)=\E \xb$.\\
The conditional error variance  $\Sigma(t)$ can be updated by the same matrix recursion as above, namely
\begin{align}
&\Sigma(t+1)= \nonumber\\ 
&= \Sigma(t)- \{\, \!\Sigma(t)\varphib(t\!+\!1)\! \left[\varphib(t\!+\!1)^{\top} \! \Sigma(t) \varphib(t\!+\!1) +\sigma^2\right]^{-1} \nonumber\\
&\q \cdot\varphib(t\!+\!1)^{\top}\!\Sigma(t) \,\}
\end{align}
with initial condition the a priori variance $\Sigma(0)=P_0$.
\end{cor}

\end{document}